\documentclass{pasj00}

\SetRunningHead{E. Akiyama, M. Momose, and H. Hayashi}{Thermal Structure of a Protoplanetary Disk around HD163296}
\usepackage{times}
\usepackage{color}

\begin{document}
\title{Thermal Structure of a Protoplanetary Disk around HD163296:\\
A Study of Vertical Temperature Distribution by CO Emission Lines}

\author{Eiji \textsc{Akiyama}, Munetake \textsc{Momose}, Hiroyuki \textsc{Hayashi}}
\affil{%
	College of Science, Ibaraki University,\\
	Bunkyo 2-1-1, Mito, Ibaraki, 310-8512, Japan.}
\email{09nd401y@mcs.ibaraki.ac.jp, momose@mx.ibaraki.ac.jp}
\and
\author{Yoshimi \textsc{Kitamura}}
\affil{Institute of Space and Astronomical Science, Japan Aerospace Exploration Agency,\\
   3-1-1, Yoshinodai, Sagamihara, Kanagawa, 229-8510, Japan.}\email{kitamura@isas.jaxa.jp}


\maketitle

\begin{abstract}
This paper presents observations of a protoplanetary disk around Herbig Ae star HD163296 in $^{12}$CO ($J$=1--0), $^{12}$CO ($J$=3--2), $^{13}$CO ($J$=1--0), and $^{13}$CO ($J$=3--2) emission lines. Double-peaked emission profiles originating from the rotating circumstellar disk were detected in all the lines. The disk parameters were estimated from model calculation in which the radial distribution of temperature or surface density inside the disk has a power-law form. The surface density should be sufficiently high so that the disk is optically thick for all the CO lines, as discussed in previous studies based on interferometric observations. The temperature and outer radius of the disk were also confirmed to be consistent with the previous results. Taking advantage of difference in position of the photosphere among the CO lines, we revealed temperature distribution in vertical direction. 
The temperature of $^{12}$CO ($J$=3--2) emitting region is about twice higher than that of any other CO emitting region; the former is $58.5\pm 9.5$ K while the latter is $31\pm15$ K at 100 AU from the central star, suggesting that there are at least two distinct temperature regions. The best fit temperature for $^{13}$CO ($J$=1--0) that should trace the deepest region of the disk is even lower, implying that there is also a different temperature region in deep inside of the disk. Such vertical temperature distribution in a disk was identified both in T Tauri and Herbig Ae stars (e.g., DM Tau, AB Aur, and HD31648), and this should be a common feature in protoplanetary disks.
\end{abstract}


\section{Introduction}
Herbig Ae (HAe) stars are widely recognized as intermediate mass counterparts of low mass T Tauri stars (TTSs) \citep{Waters98}. In the last few decades, it has been revealed that these stars are commonly accompanied by a circumstellar disk like TTSs by direct imaging at infrared and radio wavelengths \citep{Fukagawa04,Wisniewski08,ManningsSargent97,ManningsKoerner97,Pietu03}. These disks are believed to be the precursors of planetary systems around HAe stars, and are the important targets to study the formation process of a planetary system around an intermediate-mass star and to construct a general scenario for the formation of planetary systems. Recent advanced instruments have revealed their physical properties, such as the disk mass, the radial temperature and surface density distributions, the outer radius of the disk, and the velocity field. The number of well-studied disks around HAe stars, however, is still insufficient to establish a firm understanding of their physical conditions and their statistics; only a few of limited examples of their structure have deeply been analyzed (e.g., \cite{Panic08, Pietu07}). 

Temperature distribution along the vertical direction of the disk should also be taken into account when one examines how the gas and dust inside the disk evolve. Theoretical studies predict that flared circumstellar disks should have two or more layers with distinct temperature in vertical direction that are formed by diffusion of stellar radiation scattered at the disk surface \citep{Chiang97,DAlessio99,Inoue09}. To examine vertical structure of the disks based on observations, it requires examples as many as possible, though several studies have already tried to make vertical temperature or density distributions unveiled \citep{Panic08,Dartois03}. HD163296 is one of the best targets for this purpose because it exhibits strong emission from the disk in many CO isotopologue lines. HD163296 is a star with A3V in spectral type \citep{Meeus01}. Its luminosity, mass, age, and distance are 30 $L_{\odot}$, 2.3 $M_{\odot}$, 4 Myr, and 122 pc, respectively \citep{van98}. Observations at infrared and radio wavelengths revealed the existence of a circumstellar disk with its mass of 0.028 $M_{\odot}$ and  45$^{\circ}$ in inclination angle \citep{ManningsSargent97, Isella07, Hughes08}. HD163296 is categorized in group I\hspace{-.1em}I by the feature of spectral energy distribution (SED) and expected to have a self-shadowed disk \citep{Acke04,Meeus01,Dullemond04}. More recently, imaging studies with IRAM/PBI, SMA, and VLA both in continuum and in the $^{12}$CO, $^{13}$CO, and C$^{18}$O emission lines were made  \citep{Isella07}, arguing that the disk radius is 550\,AU and that the gas kinematics can be well explained by Keplerian rotation.

This paper presents multi-line observations of HD163296 with the Nobeyama 45-meter (NRO 45 m) radio telescope and the Atacama Submillimeter Telescope Experiment (ASTE) sub-millimeter telescope, and discusses vertical temperature distribution inside of the disk. The outline is as follows: Details of the observations are described in section \ref{sec:Observation}, and the results are shown in section \ref{sec:Results}. In section \ref{sec:Model Fitting}, physical properties of the disk are estimated by model fitting in which power-law distributions of temperature and surface density are assumed. Interpretation of the model fitting results as well as the comparisons with previous interferometric observations are discussed in section \ref{sec:Discussion}.

\section{Observations}
\label{sec:Observation}

\subsection{$^{12}$CO ($J$=1--0) Mapping Observations}
\label{sec:12CO(J=1-0) Mapping Observation}
Observations of $^{12}$CO ($J$=1--0) emission line were performed in winter 2006 by the NRO 45 m telescope, operated by Nobeyama Radio Observatory (NRO), a branch of National Astronomical Observatory of Japan (NAOJ). To separate the emission originated from the disk associated with the central star from surrounding components such as an envelope or a remnant cloud, a profile map including the stellar position and its vicinity with sufficient spacial resolution is decisive in identifying the origin of the emission. To obtain such a map, the multi-beam receiver, 25 Beam Array Receiver System (BEARS: \cite{Sunada00,Yamaguchi00}) that is capable of simultaneously detecting 25 different locations in double sideband (DSB) operation was employed as the frontend and tuned at the rest frequency of $^{12}$CO ($J$=1--0), 115.271 GHz. The half-power beam width (HPBW) is 15$\arcsec$ at this frequency, corresponding to 1800 AU at a distance of 122 pc from the Sun. The beam grid spacing of BEARS is 40$\arcsec$ ($\approx$ 5000\,AU) in both R.A. and Dec. directions, but this is much larger than the HPBW and therefore is insufficient to isolate the star+disk system from the ambient. To fill these gaps among the beams, we also took data at positions shifted by 20$\arcsec$ in both R.A. and Dec. directions from the central star. The integration time for the observation was 7 hours. We only show 80$\arcsec$ $\times$ 80$\arcsec$ ($\approx$ 0.05 pc $\times$ 0.05 pc) region near the star that covers a typical size of the envelope. A digital spectrometer with 32 MHz in bandwidth and 37.8 kHz in frequency resolution was used as the backend. Correction for atmospheric absorption during observations was made by the chopper-wheel method \citep{Kutner81}, and the intensities were obtained in antenna temperature, $T^*_{\rm A}$, in kelvin. Correction for the main beam efficiency, $\eta_{\rm mb}$, was made when we compared model calculation results (section \ref{sec:Model Fitting}) with the brightness temperature, $T_{\rm mb}=T^*_{\rm A}/\eta_{\rm mb}$, where $\eta_{\rm mb} =0.39$ for $^{12}$CO ($J$=1--0) emission line. The DSB system temperature in $^{12}$CO ($J$=1--0) mapping observation was between 250 K and 600 K. The accuracy for telescope pointing was regularly checked against the SiO maser VX Sgr., which was about 30$\arcmin$ away from HD163296, and pointing deviation was achieved in less than $\pm$3$\arcsec$ during whole observation period. Data reduction and analysis were made with the NEWSTAR software package developed by NRO, which is a front end of AIPS developed in the National Radio Astronomy Observatory (NRAO).

\begin{figure*}
\begin{center}
\FigureFile(180mm,120mm){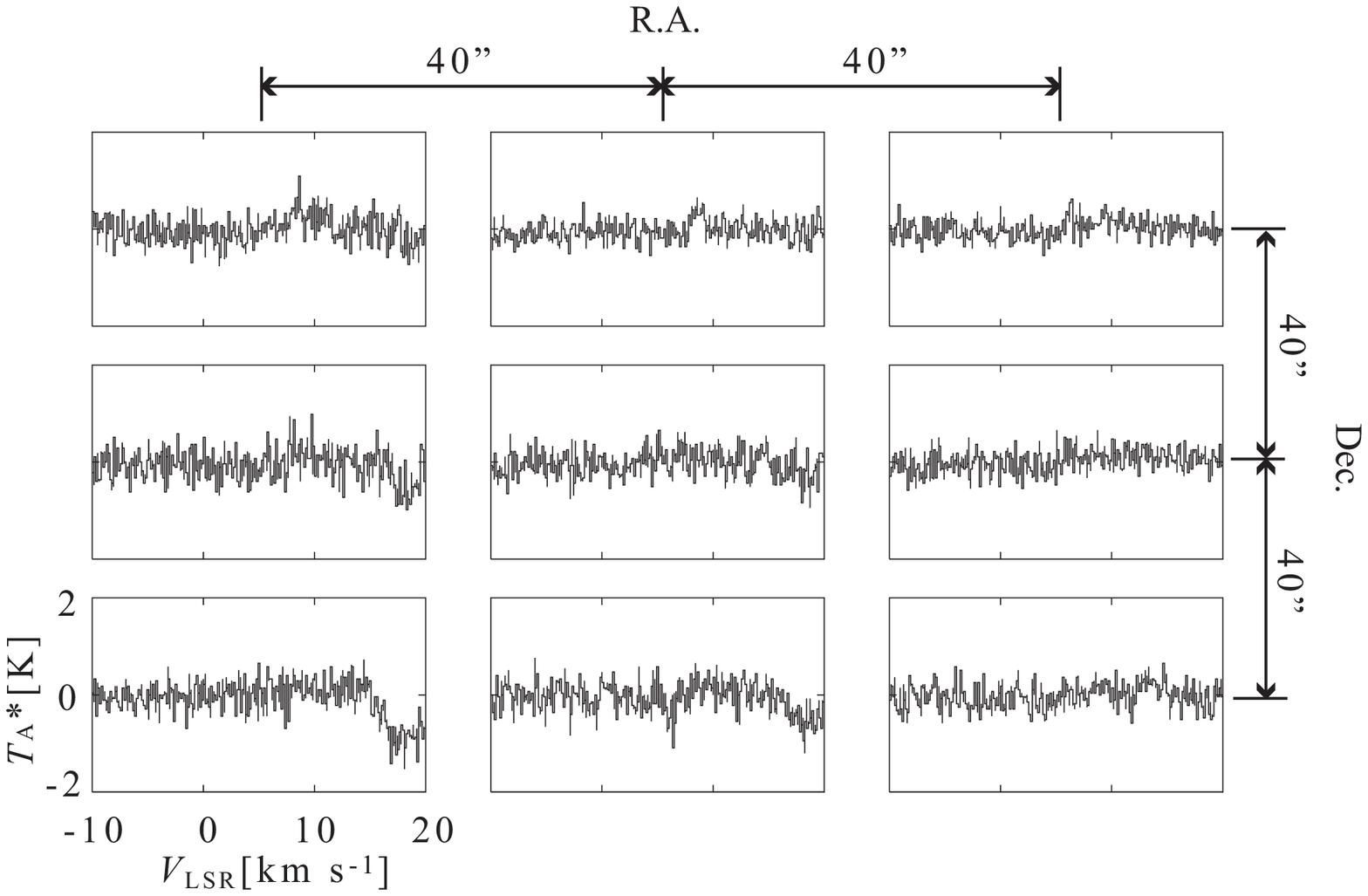}
\end{center}
\caption{Profile map of the off positions by the $^{12}$CO ($J$=1--0) observation. The central position 
is ($\alpha, \delta$)=(\timeform{17h 56m20.0s}, \timeform{-21D46m23.2s}) in J2000, and the
angular distance between adjacent points is 40$\arcsec$ in both R.A. and Dec. directions.}
\label{fig:figure1}
\end{figure*}

Observations were made in position-switching mode. It was found that the extended CO ($J$=1--0) emission was present around the region of interest, hence careful search is needed to find an off point as close to HD163296 as possible. The search for off points was conducted with BEARS based on the fact that the systemic velocity of HD163296 is 6 km s$^{-1}$ \citep{Dent05,Isella07}. The emission free position applicable as an off point was finally found at about 10$\arcmin$ north from HD163296. Figure \ref{fig:figure1} is a profile map toward this position, centered at \timeform{17h56m20.0s} and \timeform{-21D46m23.2s} (J2000) and covering 80$\arcsec$ $\times$ 80$\arcsec$ ($\approx$ 0.05 pc $\times$ 0.05 pc) region, showing that there is no prominent emission detected around 6 km s$^{-1}$ in $V_{\rm LSR}$. 

\subsection{$^{13}$CO ($J$=1--0) Emission Line Observations}
Observations of $^{13}$CO ($J$=1--0) emission line were conducted by the NRO 45 m telescope in winter 2008. As will be shown in section \ref{sec:Result of 12CO(J=1-0) Mapping Observations}, the $^{12}$CO ($J$=1--0) emission detected around 6 km s$^{-1}$ in $V_{\rm LSR}$ at the central star is isolated. Therefore, the $^{13}$CO ($J$=1--0) observations were made only at the stellar position by a pair of single-beam receivers that enable to simultaneously detect both the polarization components in single sideband (SSB) mode. The receiver was tuned at $^{13}$CO ($J$=1--0) rest frequency of 110.201 GHz and the HPBW is 15$\arcsec$ at this frequency. The $\eta_{\rm mb}$ in 2008 at this frequency was 0.4. Acousto Optical Spectrometer (AOS) whose bandwidth and frequency resolution are 40 MHz and 20 kHz respectively was used. The system noise temperature during observation was between 300 K and 600 K in SSB mode. The pointing was regularly checked by the same manner as in $^{12}$CO ($J$=1--0) observations described in section \ref{sec:12CO(J=1-0) Mapping Observation}. The same off point as in the $^{12}$CO ($J$=1--0) observations  was used. All observational data when the wind speed exceeded 5 m s$^{-1}$ were flagged out before obtaining the final profile of $^{13}$CO ($J$=1--0) emission. The resultant total integration time was 8.5 hours on source. Data reduction and analysis were made in the same manner as in $^{12}$CO ($J$=1--0) observation in section \ref{sec:12CO(J=1-0) Mapping Observation}.

\subsection{$^{12}$CO and $^{13}$CO ($J$=3--2) Observations with ASTE}
Observations of $^{12}$CO and $^{13}$CO ($J$=3--2) line were carried out in July 2006 with Atacama Submillimeter Telescope Experiment (ASTE), a 10 m sub-millimeter telescope in Chile \citep{Ezawa04}. HPBWs of the ASTE telescope at 345.79 and 330.59 GHz were 22$\arcsec$ and 23$\arcsec$, respectively, which corresponds to about $2700-2800$ AU at the distance to the target, and the $\eta_{\rm mb}$ was 0.6 around these frequencies. SIS mixer receiver having DSB response was employed. Typical atmospheric opacity at 220 GHz toward the zenith was 0.05 during the observations, and the system noise temperature of the telescope was $250-300$ K in DSB during observations. The digital auto-correlator with 1024 frequency channels was configured so that the total bandwidth was 128 MHz, resulting in a frequency resolution of 125 kHz, or $\sim$ 0.11 km s$^{-1}$ at the observed frequencies. Telescope pointing was checked every $1.5-2$ hours by cross scan of Jupiter, and the error was proved to be $\leq 3\arcsec$. All the spectra were obtained in a position-switching mode: the location of the off-position was $(\Delta\alpha, \Delta\delta) = (-20\arcmin, +20\arcmin)$ from the star. In $^{12}$CO ($J$=3--2) observations, we took spectra not only at the stellar position but also at the four adjacent points, 22$\arcsec$ apart from the stellar position in R.A. or Dec. direction, to check out whether the emission of the stellar position is really isolated. $^{13}$CO ($J$=3--2) data were taken only toward the stellar position since the $^{12}$CO ($J$=3--2) at the systemic velocity ($V_{\rm LSR} \sim 6$ km s$^{-1}$) was proven to be isolated, as will be shown in section \ref{sec:Results of ASTE Observations}. Data reduction and analysis were conducted in the same manner as in $^{12}$CO ($J$=1--0) observations described in section \ref{sec:12CO(J=1-0) Mapping Observation}.

\section{Results}
\label{sec:Results}

\subsection{$^{12}$CO ($J$=1--0) and $^{13}$CO ($J$=1--0)}
\label{sec:Result of 12CO(J=1-0) Mapping Observations}
\begin{figure*}
\begin{center}
\FigureFile(180mm,120mm){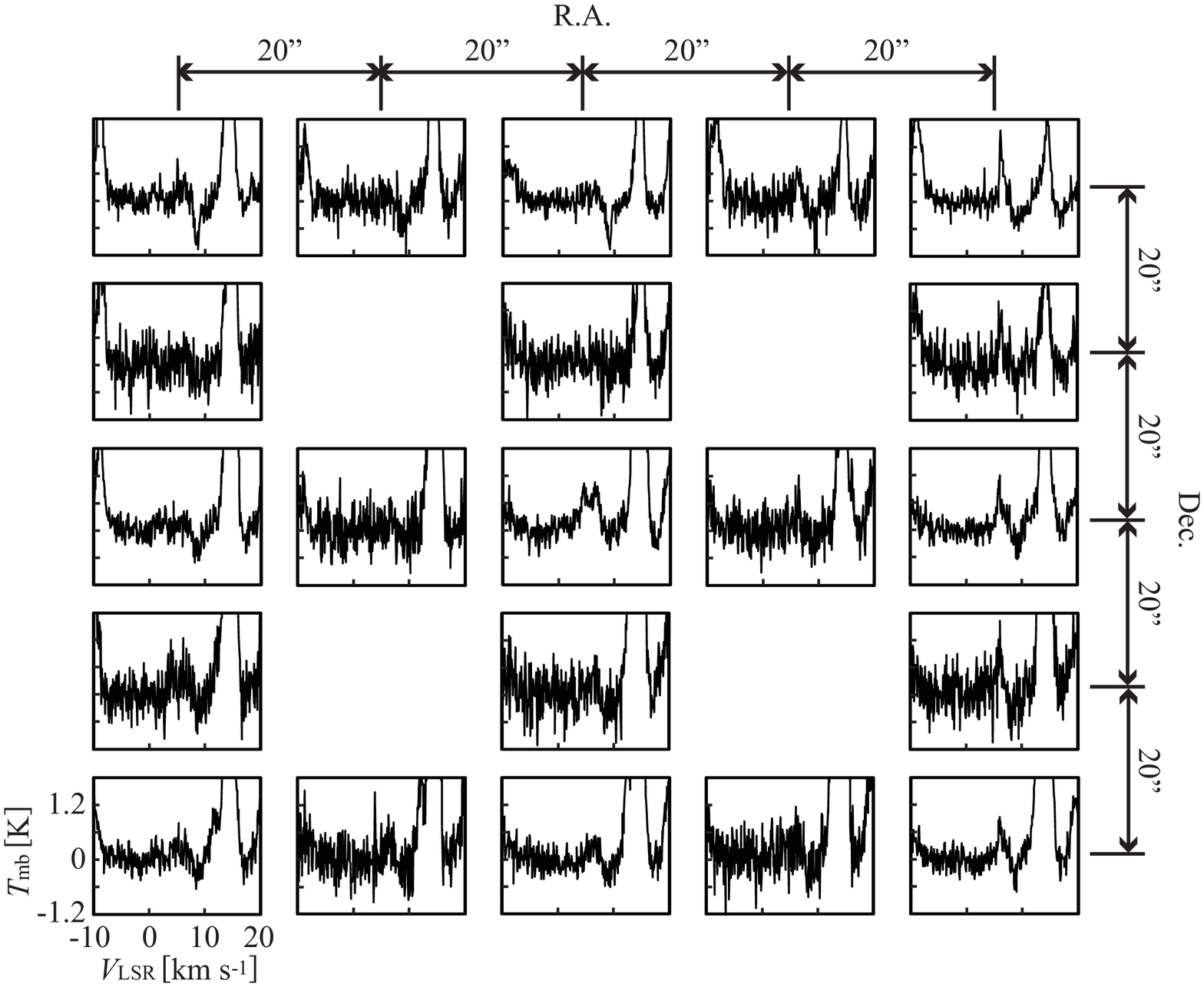}
\end{center}
\caption{Profile map of the $^{12}$CO ($J$=1--0) line around HD163296. The map center corresponds to the stellar position ($\alpha, \delta$)=(\timeform{17h 56m21.29s}, \timeform{-21D57m21.9s}) in J2000, and the angular distance between adjacent points is 20$\arcsec$ in both R.A. and Dec. directions.}
\label{fig:figure2}
\end{figure*}

\begin{figure*}
\begin{center}
\FigureFile(160mm,110mm){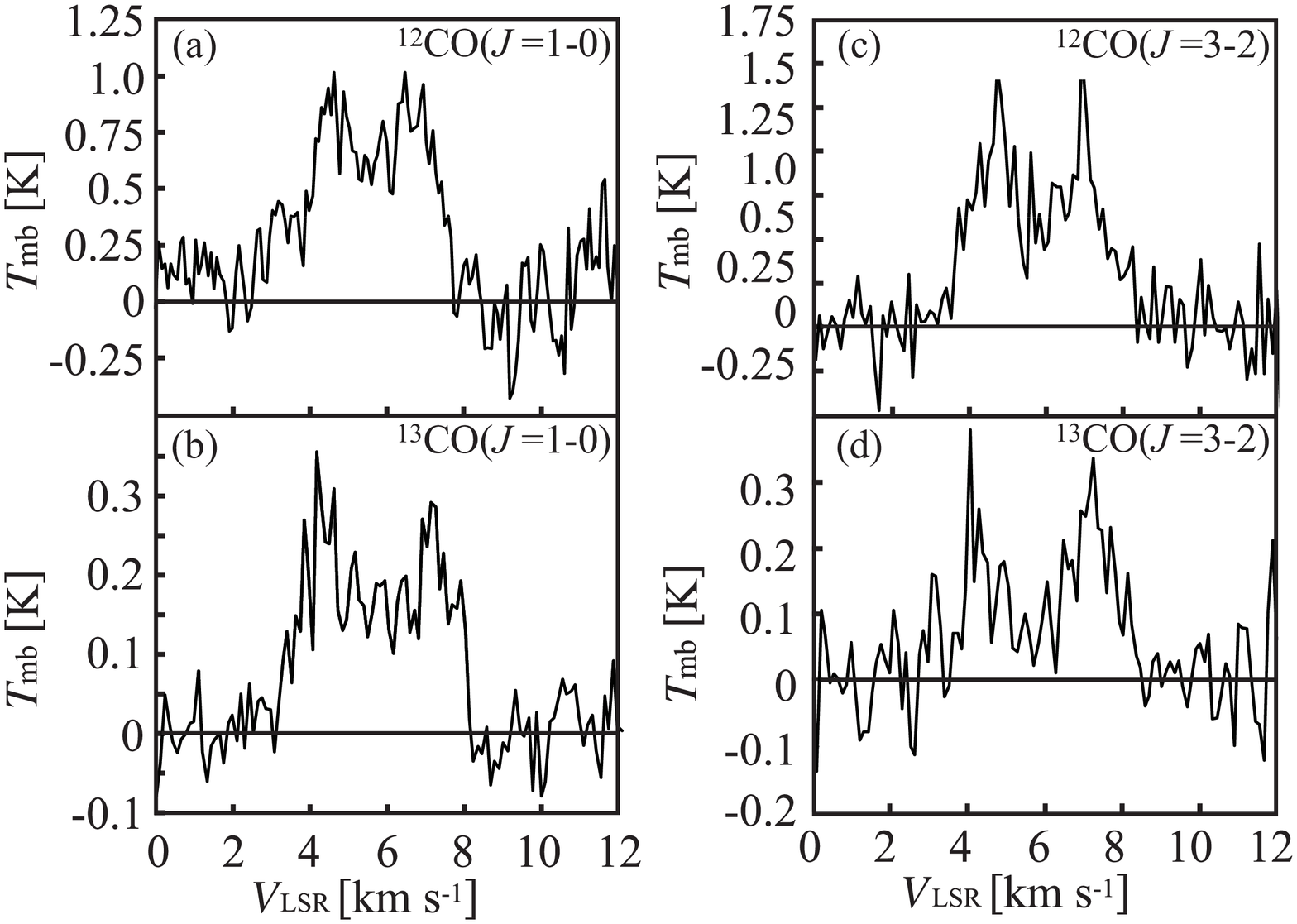}
\end{center}
\caption{Profiles detected in $^{12}$CO and $^{13}$CO  lines toward the stellar position of HD163296. 
(a) in $^{12}$CO ($J$=1--0), 
(b) in $^{13}$CO ($J$=1--0), 
(c) in $^{12}$CO ($J$=3--2) and 
(d) in $^{13}$CO ($J$=3--2).}
\label{fig:figure3}
\end{figure*}

Figure \ref{fig:figure2} shows a $^{12}$CO ($J$=1--0) profile map within the 80$\arcsec$ $\times$ 80$\arcsec$ region centered at HD163296. A clear double-peaked profile with 4.8$\sigma$ is seen only at the center panel at 6 km s$^{-1}$ in $V_{\rm LSR}$ that matches with the systemic velocity of HD163296 \citep{Dent05,Isella07}. The double-peaked profile has peak intensity of 1.0 K in $T_{\rm mb}$ and 3.2$\pm$0.1 km s$^{-1}$ in FWHM velocity width. The achieved mean rms noise level after data reduction was 0.21 K. Figure \ref{fig:figure3}a displays the same profile on the central panel in figure \ref{fig:figure2} in an enlarged view. 

\begin{figure}
\begin{center}
\FigureFile(80mm,110mm){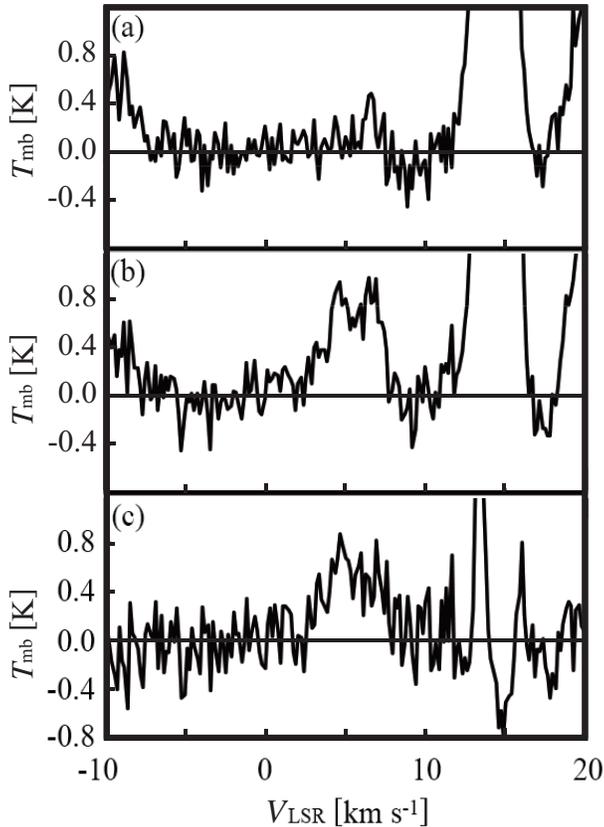}
\end{center}
\caption{(a) Averaged $^{12}$CO ($J$=1--0) profile of the four locations 20$\arcsec$ away from the central star in figure \ref{fig:figure2}. (b) The $^{12}$CO ($J$=1--0) profile toward HD163296 in figure \ref{fig:figure2}. (c) A profile after subtracting (a) from (b).}
\label{fig:figure4}
\end{figure}

Several positions in figure \ref{fig:figure2}, especially 40$\arcsec$ away to the west from the center, show an emission feature at $V_{\rm LSR}=$ 6 km s$^{-1}$, but there seems no emission above the 3$\sigma$ level at this velocity at four positions adjacent to the star (20$\arcsec$ apart from the star). To confirm whether there is no significant emission at 6 km s$^{-1}$ in these regions, these four profiles were averaged into one spectrum (figure \ref{fig:figure4}a). There is no emission above the 3$\sigma$ level at 6 km s$^{-1}$ in figure \ref{fig:figure4}a. The profile detected at the stellar position (figure \ref{fig:figure4}b, or the central panel in figure \ref{fig:figure2}) is compared with that after subtracting the profile in figure \ref{fig:figure4}a (figure \ref{fig:figure4}c). Although the noise level in figure \ref{fig:figure4}c increases due to subtraction, there is no remarkable difference between these two profiles. These comparisons strongly suggest that the double-peaked emission component at the star is isolated and originates from the compact circumstellar disk in Keplerian rotation. The components at $V_{\rm LSR}=$ 6 km s$^{-1}$ in south and west regions $\ge 40\arcsec$ apart from the star in figure \ref{fig:figure2} might be the emissions from a remnant cloud partially remained around the star+disk system.

The dip at 8 km s$^{-1}$ around the star+disk system (figure \ref{fig:figure2}) is due to weak emission seen at the off points, especially in the northern region (figure \ref{fig:figure1}); in position switching mode, the emission at the off point appears like an absorption dip. The dip dose not affect further analysis since it does not corrupt the original emission profile at 6 km s$^{-1}$. In addition, the strong emission components between 10 and 15 km s$^{-1}$ and at 20 km s$^{-1}$ appear regardless of any positions: these two components probably come from foreground or background clouds that exist coincidentally along a line of sight.

Figure \ref{fig:figure3}b shows the profile of $^{13}$CO ($J$=1--0) emission line. Again, a double-peaked profile was clearly detected. The peak intensity and velocity width are 0.36 K in $T_{\rm mb}$ and 3.8$\pm$0.2\,km s$^{-1}$ at FWHM, respectively. The achieved rms noise level is 0.061 K. Since the $^{12}$CO ($J$=1--0) and $^{13}$CO ($J$=1--0) emissions are detected in the same velocity range, both emissions should have the same origin. The intensity ratio between $^{12}$CO($J$=1--0) and $^{13}$CO($J$=1--0) is less than 5, suggesting that the disk is optically thick. \citet{Isella07} evaluated the optical depth of the CO emission lines and concluded that the disk is optically thick. Our model fitting shown in section \ref{sec:Fitting Results} also provided that the disk is optically thick for both the CO lines.

\subsection{$^{12}$CO ($J$=3--2) and $^{13}$CO ($J$=3--2)}
\label{sec:Results of ASTE Observations}
\begin{figure*}
\begin{center}
\FigureFile(180mm,120mm){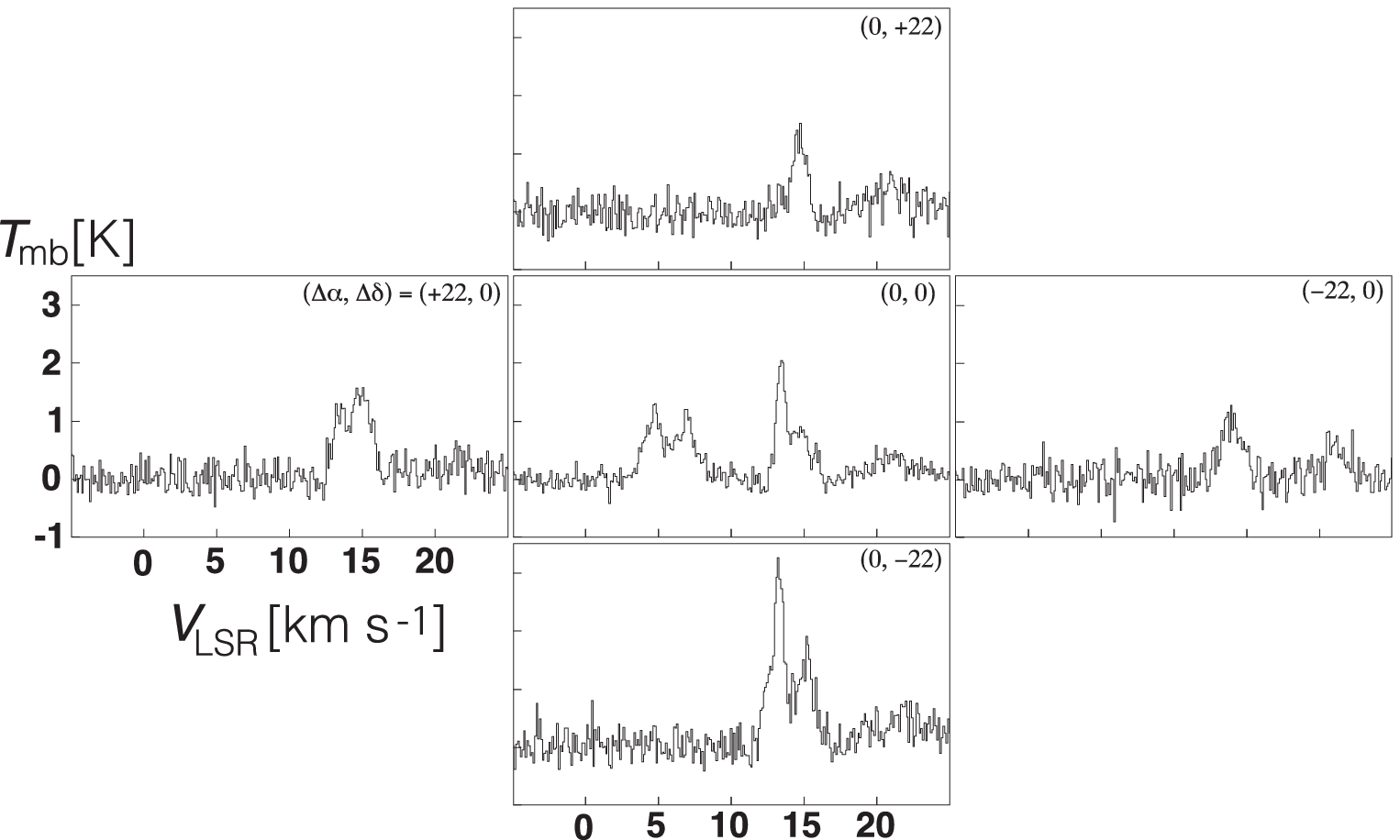}
\end{center}
\caption{Profile map of $^{12}$CO ($J$=3--2) around HD163296. The offset (in arc-second) from the stellar position is shown in the top right corner of each panel.}
\label{fig:figure5}
\end{figure*}

Figure \ref{fig:figure5} shows a $^{12}$CO ($J$=3--2) profile map within the 44$\arcsec$ $\times$ 44$\arcsec$ region centered at HD163296; the central panel of figure \ref{fig:figure5} is also shown in figure \ref{fig:figure3}c in an enlarged view. The achieved rms level was 0.122 K. A clear double-peaked profile at $\sim $6 km s$^{-1}$ is seen only in the center panel. The peak intensity and velocity width are 1.35 K in $T_{\rm mb}$ and 3.1$\pm$0.1 km s$^{-1}$ in FWHM, respectively. The emission components between 10 and 15 km s$^{-1}$ and at $\sim$20 km s$^{-1}$ in $V_{\rm LSR}$ appear regardless of positions as in the case of $^{12}$CO ($J$=1--0) (figure \ref{fig:figure2}).

Figure \ref{fig:figure3}d presents the $^{13}$CO ($J$=3--2) spectrum at the stellar position, showing its double-peaked profile. The peak intensity and velocity width are 0.38 K in $T_{\rm mb}$ and 3.85 km s$^{-1}$ in FWHM, respectively. As in the case of the $J$=1--0 lines, the disk seems optically thick for $J$=3--2 line from the intensity ratio between $^{12}$CO ($J$=3--2) and $^{13}$CO ($J$=3--2)(see also section \ref{sec:Fitting Results}). 

\section{Model Fitting}
\label{sec:Model Fitting}
Model fitting is one of the effective methods to estimate physical properties of disks around HAe/Be stars \citep{Dullemond01,Dullemond02,Testi03,Natta04}. We fitted all the observational profiles, which are two CO isotopologues' emission lines with two different rotational transitions, with those calculated by the disk model (details in section \ref{sec:Disk Model}). We adopted a classical disk model which includes power-law forms of temperature and surface density distributions in radial direction with inner and outer cutoffs \citep{Kitamura93}. The results are compared with the ones obtained by millimeter and sub-millimeter interferometers (\cite{Isella07}, details in section \ref{sec:interferometers}). Although the interferometric observations can provide detailed geometrical structure with high spacial resolution, it might be unable to detect the emission from the diffuse gas uniformly extending away due to spatial filtering effect. The single dish telescopes with large diameter, on the other hand, have higher sensitivity to the weak extended emission which interferometers can not detect well. Interferometric and single dish observations are, therefore, complementary to each other, and comparing their results is a fruitful way to gain a better understanding of the disk nature.

\subsection{Disk Model}
\label{sec:Disk Model}
Several assumptions for the disk structure and parameters are applied in the model fitting. First, the disk is assumed to have Keplerian rotation as was proven to be the cases in other HAe stars such as HD169142 and HD31648 \citep{Panic08,ManningsKoerner97}. \citet{Isella07} pointed out the existence of Keplerian rotating disk around HD163296 from ``a butterfly shape'' seen in its position velocity diagram. Secondly, the temperature distribution over the disk, $T(r)$, is assumed to follow a power-law form of the radius with its index of $-0.5$, as described by the following equation:     

\begin{equation}
T(r)=T_0\left(\frac{r}{100\rm{AU}}\right)^{-0.5},
\label{eq:temperature}
\end{equation}
where $r$ is the radius in cylindrical coordinate system centered on the star and $T_{\rm 0}$ is the temperature at $r=100$ AU \citep{Hayashi85,Beckwith90}. This assumption implies that the temperature is constant along the vertical direction ($z$) at a certain $r$. In reality, a circumstellar disk should have temperature distribution along the $z$-direction. An elaborate disk model (e.g., \cite{Chiang97,Tanaka05}) predicts that the surface regions where the stellar radiation can penetrate and heat directly will have higher temperature than that in the interior regions, forming 2-layer temperature structure. Single layer temperature model, however, can still be applicable because an optically thick line should have ``photosphere'' at each $r$ and its characteristic temperature can uniquely be determined. That is, vertical temperature distribution can be traced by a series of a different temperature of each CO emission line. 

Thirdly, the radial distribution of surface density over the disk, $\Sigma(r)$, is also assumed to follow a power-law form of the radius with its index of $-1.0$ \citep{Isella07}, as described by the following equation:  

\begin{equation}
\Sigma(r)=\Sigma_0\left(\frac{r}{100\rm{AU}}\right)^{-1.0},
\label{eq:sigma}
\end{equation}
where $\Sigma_{\rm 0}$ is the surface density at $r=100$ AU. Fourthly, we assumed that density distribution in vertical direction can be achieved by hydrostatic equilibrium. The density distribution $\rho(r,z)$ is, therefore, expressed by 

\begin{equation}
\rho(r,z)=\rho(r,0){\rm exp}\bigg[-\left(\frac{z}{H(r)}\right)^2\biggr]\ ,
\label{eq:rho}
\end{equation}
where $\rho(r,0)$ is the density at the mid-plane of the disk, and $H(r)$ is the scale height given by

\begin{equation}
H(r)=\sqrt{\frac{2r^3k_{\rm B}T(r)}{GM_\ast m}}\ ,
\label{eq:scale_height}
\end{equation}
where $k_{\rm B}$ is Boltzmann constant, $G$ is gravitational constant, $M_\ast$ is the mass of the central star, and $m$ is the mean mass of gas molecules. Thus, $\rho(r,0)$ in equation (\ref{eq:rho}) becomes
\begin{equation}
\rho(r,0)=\frac{\Sigma (r)}{\sqrt{\pi}H(r)}\ .
\end{equation}

Fifthly, local thermodynamic equilibrium (LTE) is assumed. The number density of H$_2$ at the position of photosphere of each CO line can be estimated by the equation (\ref{eq:rho}). As shown in section \ref{sec:Fitting Results}, the number density of H$_2$ at the $^{12}$CO ($J$=3--2) photosphere is calculated to be $\sim$ 10$^{7}$ cm$^{-3}$ at $r=$ 100 AU, and it is higher than the typical value of critical density for $^{12}$CO ($J$=3--2) ($\sim$ 5 $\times$ 10$^{4}$ cm$^{-3}$ at 60 K, fitting temperature of this line discussed in section \ref{sec:Fitting Results}) by more than two orders of magnitude. LTE assumption is sometimes inadequate for analysis of protoplanetary disks since it tends to overestimate the excitation temperature for a higher transition with a high critical density. However, \citet{Pav07} evaluated many types of model calculations and found that LTE approximation seems valid for the cases of $J$=4--3 or a lower transition of CO. The LTE assumption is, therefore, applicable in our model calculation for the $J$=1--0 and $J$=3--2 transitions of $^{12}$CO and $^{13}$CO. 

Finally, the abundance ratios of H$_{2}$ to $^{12}$CO and $^{12}$CO to $^{13}$CO are assumed to be 10$^4$ and 60, respectively, which are the typical values in the interstellar medium \citep{Frerking82,Wannier82}. The disk is assumed to be heated only by a stellar radiation and accretion heating is neglected. This is a reasonable approximation for a disk with modest accretion rate such as HD163296 case (7.6 $\times$ 10$^{-8}$ $M_{\odot}$ yr$^{-1}$; \cite{Garcia06}).
 
All parameters regarding the object, listed in table \ref{tb:parameters}, should be taken into account in the model calculation for the best accuracy, but some of them rarely affect the disk property. Emission profiles are mostly characterized by three parameters, $T_{\rm 0}$, $\Sigma_{\rm 0}$, and the outer radius of the disk ($r_{\rm out}$), and these are regarded as variables. The other parameters are fixed to be constants as listed in table \ref{tb:parameters}. 

In actual calculations, $\Sigma_{\rm 0}$ was fixed in each run, and the possible fitting range of $T_{\rm 0}$ was searched. $\Sigma_{\rm 0}$ was set to be between 0.003 and 3 g cm$^{-2}$. The search for $T_{\rm 0}$ in upper range of $\Sigma_{\rm 0}$ is stopped at 3 g cm$^{-2}$ because $T_{\rm 0}$ in each CO line is well confined due to asymptotic behavior of fitting temperature in large $\Sigma_{\rm 0}$ region and the result remains almost the same. Intensity is affected by both $T_{\rm 0}$ and $\Sigma_{\rm 0}$ in the case of optically thin disk and by only $T_{\rm 0}$ in the case of optically thick disk. Among three variables, $r_{\rm out}$ can be guessed from the separation between two peaks in the profiles obtained from observation, since the Keplerian velocity at $r_{\rm out}$ ($V(r_{\rm out})$) roughly corresponds to half of the separation between the two peaks of the profile. $r_{\rm out}=$1000 AU is adopted as an initial guess and then adjusted.

\subsection{Fitting Results}
\label{sec:Fitting Results}

\begin{figure*}
\begin{center}
\FigureFile(180mm,80mm){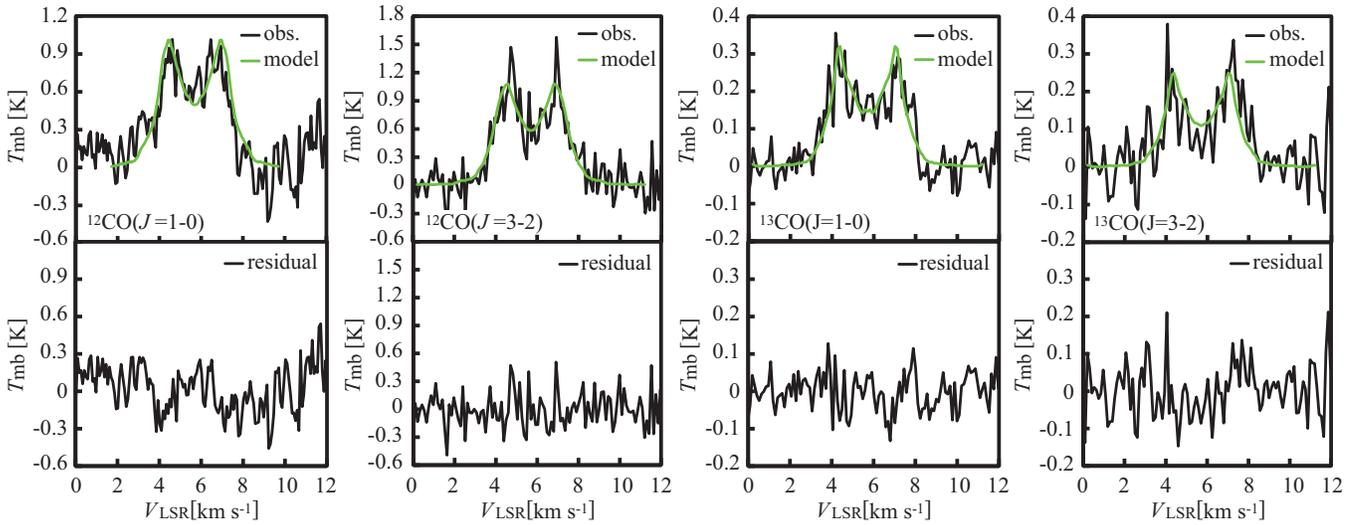}
\end{center}
\caption{(Upper panels) Comparisons between the line profiles obtained by our observations (solid line) and best-fit model calculations when $\Sigma _0 = 0.1$ g cm$^{-2}$ (green lines). Best-fit values for $T_0$ and $r_{\mathrm{out}}$ in the model calculations are listed in table \ref{tb:best fit parameters}. (Lower panels) Difference between the observed and calculated profiles shown in the upper panels.}
\label{fig:figure6}
\end{figure*}

\begin{figure*}
\begin{center}
\FigureFile(170mm,150mm){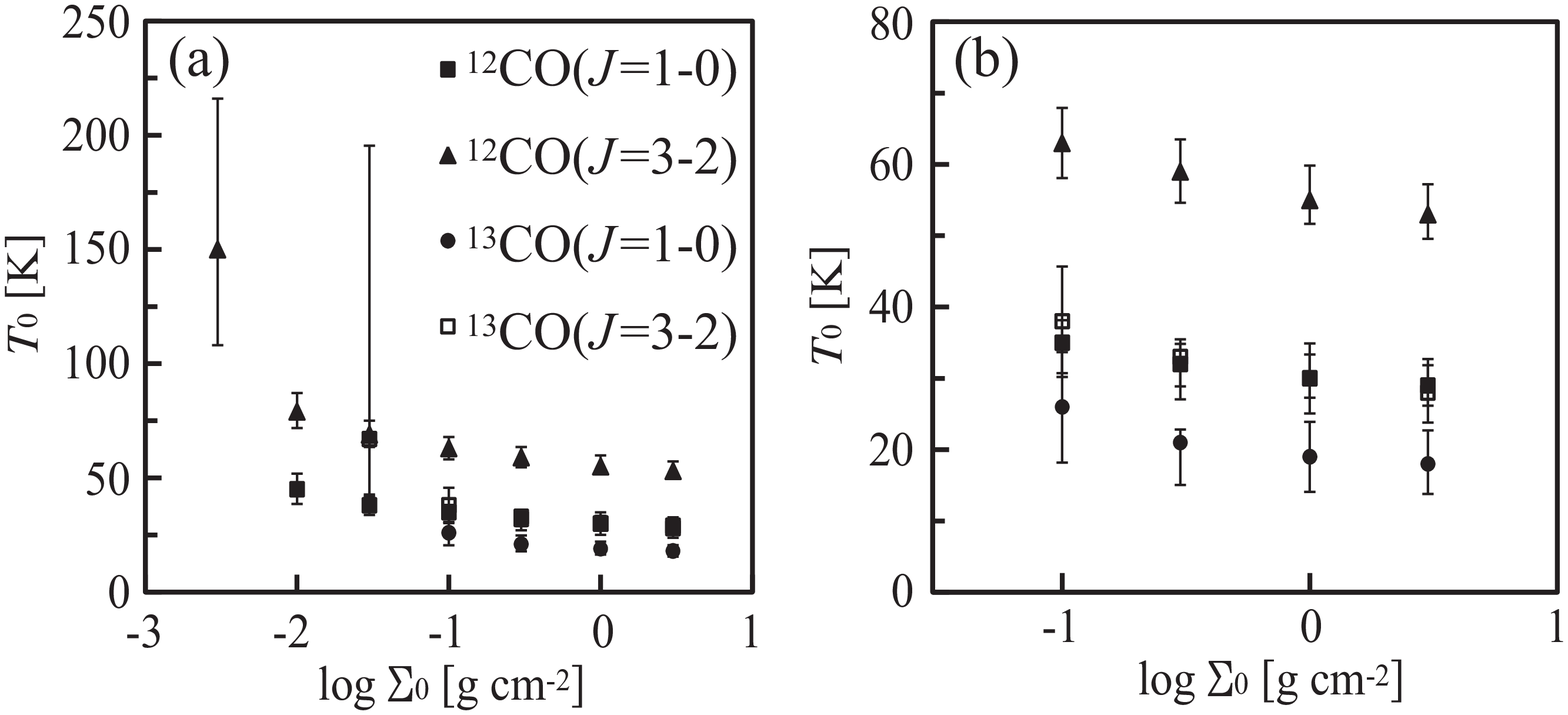}
\end{center}
\caption{Best-fit values of $T_0$ with uncertainties for the acceptable fit in model calculations as a function of $\Sigma_{\rm 0}$. The panel (a) shows the cases of $\Sigma_{\rm 0} = 0.003-3$ g cm$^{-2}$, and the panel (b) magnifies the lower right part of (a).}
\label{fig:figure7}
\end{figure*}

The best fit temperatures and outer radius at 0.1, 0.3, 1, and 3 g cm$^{-2}$ in $\Sigma_{\rm 0}$ where fitting solution for all the CO lines simultaneously exist are provided in table \ref{tb:best fit parameters}, and the fitting results are shown in figure \ref{fig:figure6}. Note that the observed profiles shown in figure \ref{fig:figure6} were re-sampled with velocity resolution of 0.08 km s$^{-1}$ for $^{12}$CO ($J$=1--0) line and 0.11 km s$^{-1}$ for the others to directly compare with the model calculations. As shown in table \ref{tb:best fit parameters}, the fitting temperature of $^{12}$CO ($J$=3--2) shows the highest value among all the CO lines. It is essential to ascertain the goodness of the solution obtained in the model fitting. Figure \ref{fig:figure7} shows fitting temperature at several specific $\Sigma_{\rm 0}$ between 0.003 and 3 g cm$^{-2}$. We regarded the results as an acceptable fit when the following two conditions were met simultaneously: (i) to exclude the case of large deviation in amplitude of profiles, the residual in absolute value integrated over the range of $T_{\rm mb}\geq$ 2$\sigma$ should be less than the uncertainty (1$\sigma$) for the integrated intensity, (ii) to exclude the case of large deviation in line shape, more than 80\% of the channels over $T_{\rm mb} > 3\sigma$ should agree with the model calculations within 2$\sigma$. 

The best fit $T_0$ for the $^{12}$CO ($J$=3--2) line in table \ref{tb:best fit parameters} are between 58.5 $\pm$ 9.5 K, and this range is significantly higher than those estimated for the other CO lines, 31 $\pm$ 15 K when $\Sigma_{\rm 0}$ is greater than 0.1 g cm$^{-2}$. This difference in best fit $T_0$ implies that the disk is composed of at least two layers with distinct temperatures. The uncertainty of fitting temperature is smaller in higher surface density or large optical depth in the disk. In lower surface density, the fitting temperature range becomes drastically larger or no solution exists; $^{13}$CO ($J$=1--0) emission line, for instance, dose not have any fitting temperature at $\Sigma_{\rm 0}$ less than 0.1 g cm$^{-2}$. The fitting errors in temperature at 0.1 g cm$^{-2}$ in $\Sigma_{\rm 0}$ are more than 30 \% for $^{13}$CO ($J$=1--0) line and more than 20 \% for $^{13}$CO ($J$=3--2) line, respectively. The ones at greater than 0.3 g cm$^{-2}$ in $\Sigma_{\rm 0}$, on the other hand, are less than 20 \% for all the CO lines and the fitting temperatures are gradually confined. The $\Sigma_{\rm 0}$ that is greater than 0.1 g cm$^{-2}$ obtained by model fitting corresponds to $\gtsim$ 10 ($\sim$ 5 even at $r_{\rm out}$ = 550 AU) in optical depth, and therefore the disk is surely optically thick for all the CO lines.  

\begin{table*}
\begin{center}
\caption{Parameters in model calculation}
\begin{tabular}{ccc}
\hline
\multicolumn{2}{c}{Invariables} & \ \ \ \ Variables\ \ \ \\ \hline \hline
$M_{\rm \ast}$ &\ \ 2.3 $\MO$\footnotemark[$*$] & $T_{\rm 0}$ \\
$d$ &\ 122 pc\footnotemark[$\ast$] &$\Sigma_{\rm 0}$ \\
$\theta$ &\ 45$^\circ$\footnotemark[$\dagger$$\ddagger$] &$r_{\rm out}$ \\
$r_{\rm in}$ &\ 0.1\,AU & \\
$p$ &\ 1.0\footnotemark[$\dagger$$\ddagger$] & \\
$q$ &\ 0.5 & \\ 
$X{\rm (^{12}CO)}$ &\ 1/10000 & \\ 
$X{\rm (^{12}CO)}/X{\rm (^{13}CO)}$ &\ 60 & \\ 
\hline
\multicolumn{2}{@{}l@{}}
{\hbox to 0pt{\parbox{85mm}{\footnotesize
\par\noindent
\footnotemark[$\ast$] van den Ancker, de Winter, \& Djie (1998).
\par\noindent
\footnotemark[$\dagger$] Isella et al. (2007)
\par\noindent
\footnotemark[$\ddagger$] Hughes et al. (2008)
}\hss}}
\end{tabular}
\label{tb:parameters} 
\end{center}
\end{table*} 

\begin{table*}
\begin{center}
\caption{Best Fit Parameters at $r_{\rm out}=$ 550AU}
\begin{tabular}{ccccc}
\hline
\multicolumn{1}{c|}{} & \multicolumn{4}{c} {$T_0$ [K]}\\
\multicolumn{1}{c|}{$\Sigma _0$ [g cm$^{-2}$]} & $^{12}$CO ($J$=3--2) & $^{13}$CO ($J$=3--2) & $^{12}$CO ($J$=1--0) & $^{13}$CO ($J$=1--0) \\ \hline \hline
\multicolumn{1}{c|}{0.1} & 63$^{+5}_{-5}$ & 38$^{+8}_{-8}$ & 35$^{+3}_{-4}$ & 26$^{+8}_{-6}$ \\ 
\multicolumn{1}{c|}{0.3} & 59$^{+4}_{-3}$ & 33$^{+2}_{-6}$ & 32$^{+4}_{-3}$ & 21$^{+4}_{-3}$ \\
\multicolumn{1}{c|}{1} & 55$^{+5}_{-3}$ & 30$^{+5}_{-5}$ & 30$^{+3}_{-3}$ & 19$^{+3}_{-3}$ \\
\multicolumn{1}{c|}{3} & 53$^{+4}_{-4}$ & 28$^{+5}_{-4}$ & 29$^{+3}_{-3}$ & 18$^{+3}_{-2}$ \\
\hline
\end{tabular}
\label{tb:best fit parameters}
\end{center}
\end{table*}

\section{Discussion}
\label{sec:Discussion}

\subsection{Comparison with observation by interferometers}
\label{sec:interferometers}
High-resolution images of HD 163296 were taken with interferometers at multi-wavelengths (the IRAM/PBI, SMA, and VLA) both in continuum and in the emission lines of $^{12}$CO, $^{13}$CO, and C$^{18}$O \citep{Isella07}. They compared the observational results of the CO emissions with those predicted by self-consistent disk model to estimate physical disk properties. Although the assumptions in model calculations were slightly different from each other, the temperature and minimum surface density of the disk derived in section \ref{sec:Fitting Results} showed a reasonable agreement with the values obtained by \citet{Isella07}; the temperature at 100 AU and the minimum surface density at 10 AU from the central star were estimated to be 26$^{+8}_{-6}$ K and 1 g cm$^{-2}$ from our $^{13}$CO ($J$=1--0) observations and they were within the possible range of the fitting values obtained by \citet{Isella07}: 30$\pm10$ K and 4$^{+12}_{-3}$ g cm$^{-2}$, respectively. The radius derived from single dish observations, which generally have higher sensitivity to weaker emission, was 550$\pm100$ AU and was consistent with the one derived by interferometric observations.

\subsection{Thermal Structure of the Disk}
As described in section \ref{sec:Fitting Results}, the disk is arguably optically thick in CO emissions. Taking advantage of different mass opacity of isotopologues whose emission lines reflect the temperature at each photosphere having different depth, the vertical thermal structure of the disk can be analyzed. As shown in figure \ref{fig:figure7}, model fittings in the case of $^{12}$CO ($J$=1--0), $^{13}$CO ($J$=1--0), and $^{13}$CO ($J$=3--2) emission lines were succeeded in almost the same temperature when $\Sigma_{\rm 0}>$ 0.1 g cm$^{-2}$, but the $^{12}$CO ($J$=3--2) results require almost double of the fitting temperature of the other three lines. These different temperatures can be explained by temperature distribution along the vertical direction, as predicted by theoretical studies (e.g., \cite{Chiang97}). 

To examine the vertical temperature distribution in detail, we derived the column densities of H$_{2}$ that satisfy optical depth for each CO emission line to be unity ($N_{\tau_{\rm \nu_0}=1}$) from 
\begin{eqnarray}
N_{\tau_{\nu_0}=1}=\frac{3{k_B}T_{\rm ex}\Delta{V_{\rm gas}}}{8{\pi}^3B\mu^2{\rm cos}\theta}\frac{1}{(J+1)}{\rm exp}\left[\frac{hBJ(J+1)}{k_BT_{\rm ex}}\right]\nonumber\\
\times \left[1-{\rm exp}\left(\frac{h{\nu_0}}{{k_B}T_{\rm ex}}\right)\right]^{-1},
\label{eq:N}
\end{eqnarray} 
where $B$ is the rotational constant of a molecule, $\mu$ is the permanent dipole moment, $T_{\rm ex}$ is the excitation temperature, $\Delta{V_{\rm gas}}$ is the velocity width of a gas molecule, $\theta$ is the inclination angle, $h$ is Planck constant, $J$ is a rotational quantum number, and $\nu_{\rm 0}$ is the frequency of transition \citep{Scoville86}. Note that $N_{\tau_{\rm \nu_0}=1}$ is the value obtained by integrating the number density from the disk surface to the disk interior along a line of sight.

In this paper, we adopted $\Delta{V_{\rm gas}}=\Delta{V_{\rm th}}$ for simplicity, where $\Delta{V_{\rm th}}$ is the thermal velocity width of gas molecules expressed by 

\begin{equation}
\Delta{V_{\rm gas}}\equiv\Delta{V_{\rm th}}=\sqrt{\frac{8{\rm ln}2k_{\rm B}T(r)}{m_{\rm CO}}}.
\label{eq:thermal velocity}
\end{equation}
Turbulent velocity ($V_{\rm turb}$) can also contribute to $\Delta{V_{\rm gas}}$. \citet{Hughes11} derived a turbulent line width of $\sim$ 0.3 km s$^{-1}$ for the disk around HD163296 by fitting profile with high spectral resolution. Previous observations also revealed that $V_{\rm turb}$ in disks around TTS or HAe star is comparable to $\Delta{V_{\rm th}}$: 0.07-0.15 km s$^{-1}$ in DM Tau \citep{Dartois03,Simon01} and 0.38 km s$^{-1}$ in AB Aur \citep{Pietu05}. In such cases, however, the line profile calculated by the model is not altered significantly by $\Delta{V_{\rm turb}}$ because the line shape is mainly determined by Keplerian rotation velocity. This assumption probably produces a factor of uncertainty for the estimation of $N_{\tau_{\rm \nu_0}=1}$ but it does not significantly affect the following discussion.

\begin{figure*}[b!]
\begin{center}
\FigureFile(170mm,150mm){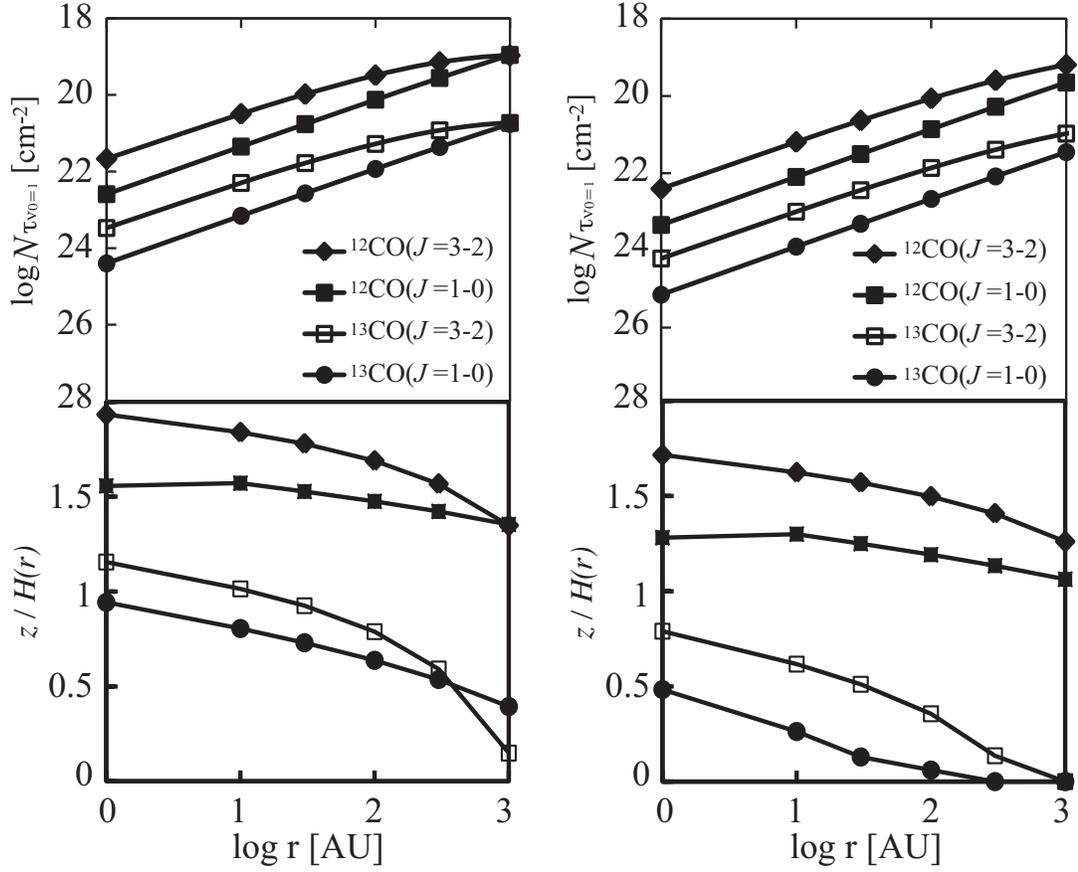}
\end{center}
\caption{(upper panels) Column densities in $N(\mathrm{H}_2)$ that are required to become $\tau_{\rm \nu_0}=1$ for each CO emission line. The radial temperature dependence is set to be $T(r) = 30 (r/100\mathrm{AU})^{-0.5}$ [K] (left) and $T(r) = 60 (r/100\mathrm{AU})^{-0.5}$ [K] (right), respectively. Note that the vertical axises are reversed. (lower panels) Vertical distance from the mid-plane of the disk normalized by the scale height of the local temperature in the case of $\Sigma_{\rm 0} =$ 0.1 g cm$^{-2}$. These vertical locations of each CO line correspond to the location corresponding to $N_{\tau_{\rm \nu0=1}}$ shown in the upper panel.} 
\label{fig:figure8}
\end{figure*}

$N_{\tau_{\rm \nu_0}=1}$ for each CO emission line provides the information about the vertical location of the photosphere of each CO emission line in the disk. Figure \ref{fig:figure8} shows $N_{\tau_{\rm \nu_0}=1}$ and vertical locations of photospheres of each CO emission line. The results are provided in the case of $T_0=30$ K and 60 K, which are the representative values of the low temperature interior and the high temperature surface, respectively, as described in section \ref{sec:Fitting Results}. To derive the location of each photosphere, on the other hand, we select $\Sigma_{\rm 0} =$ 0.1 g cm$^{-2}$ as the representative case since this value is consistent with the results by \citet{Isella07}. In both cases shown by figure \ref{fig:figure8}, $N_{\tau_{\rm \nu_0}=1}$ results in the descending order of $^{12}$CO ($J$=3--2), $^{12}$CO ($J$=1--0), $^{13}$CO ($J$=3--2) and $^{13}$CO ($J$=1--0), implying that the photosphere of $^{12}$CO ($J$=3--2) emission is located uppermost among the four CO lines. The total geometric cross section of grains per hydrogen atom ($\sigma_{\rm d}$) in interstellar medium is $\sim$10$^{-21}$ cm$^{2}$ at visible wavelength \citep{Stahler04}, hence the column density that makes the optical depth for the stellar radiation unity is 10$^{21}$ cm$^{-2}$. The grazing angle is assumed to be 0.05 in \citet{Inoue09} calculation, hence the column density of the upper layer should be about 5 $\times$ 10$^{19}$ cm$^{-2}$. This column density is consistent with the one above the photosphere of $^{12}$CO ($J$=3--2) at $r\gtsim 100$ AU, as shown in figure \ref{fig:figure8}. $N_{\tau_{\rm \nu_0}=1}$ of $^{12}$CO ($J$=1--0) decreases faster than that of $^{12}$CO ($J$=3--2) at $r >300$ AU in the case of $T_{0}$ = 30 K because CO molecules is not sufficiently excited to the $J$=3 level due to lower temperatures. The same phenomenon is seen in $^{13}$CO as well. From the comparisons between the fitting temperature and the vertical locations of the photosphere of each CO line, the general trend is that the farther away the layer is from the mid-plane of the disk, the warmer it reaches. This is quite reasonable because the gas around the disk surface should be heated more by the radiation from the central star. 

\begin{figure*}[t]
\begin{center}
\FigureFile(90mm,100mm){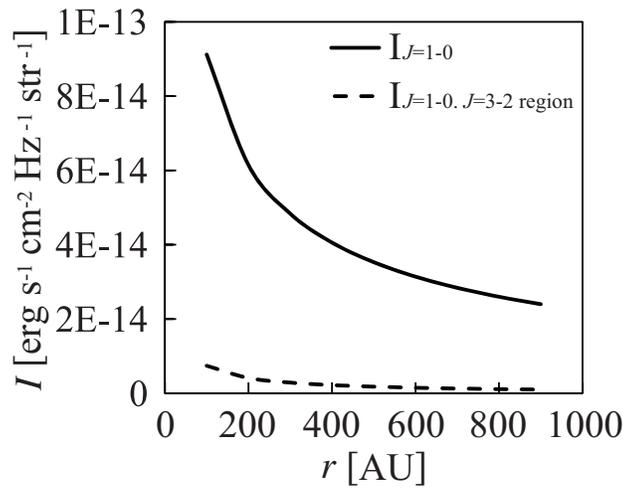}
\end{center}
\caption{Influence of the higher temperature layer on the CO ($J$=1--0) emission by higher temperature layer. The solid line represents the intensity of $^{12}$CO ($J$=1--0) emission at the fitting temperature and surface density ($T_{\rm 0}$ = 30 K and $\Sigma_{\rm 0}$ = 0.1 g cm$^{-2}$). The dashed line represents the intensity of $^{12}$CO ($J$=1--0) emission line from the region where the optical depth of $^{12}$CO ($J$=3--2) is less than unity when $T_{\rm 0}$ = 60 K.}
\label{fig:figure9}
\end{figure*}

When we derive $N_{\tau_{\rm \nu_0}=1}$ in figure \ref{fig:figure8}, we assumed that the disk temperature is uniform in vertical direction. These considerations, however, might be invalid as a consequence of vertical temperature distribution; the $^{12}$CO ($J$=3--2) emission line originates from an upper layer of the disk, but this could also contribute significantly to the other CO lines. We therefore evaluate how strongly such an upper layer affects the other CO emissions. Figure \ref{fig:figure9} shows how strongly the $^{12}$CO ($J$=1--0) is emitted from above the photosphere of $^{12}$CO ($J$=3--2). In this figure, we first calculate the optical depth of $^{12}$CO ($J$=1--0) emission line in the region above the photosphere of $^{12}$CO ($J$=3--2), and then the intensity of $^{12}$CO ($J$=1--0) at the fitting temperature was compared with that originated from the region. The contribution of high temperature layer is less than 10$\%$ of the $^{12}$CO ($J$=1--0) intensity and can be safely neglected. Therefore, the estimated vertical location of each CO emission layer in figure \ref{fig:figure8} is validated. 

As shown in figure \ref{fig:figure7}, the temperature of $^{12}$CO ($J$=1--0) and $^{13}$CO ($J$=3--2) layers are almost the same when $\Sigma_{\rm 0}>$ 0.1 g cm$^{-2}$. The temperature of $^{13}$CO ($J$=1--0) layer seems to be even lower. Such vertical thermal structure of irradiated accretion disks around TTS has been theoretically predicted. \citet{DAlessio99} constructed the detailed vertical structure models of irradiated accretion disks and extensively explored the dependence of the structure and emission properties on mass accretion rate, viscosity parameter, and disk radius. \citet{Inoue09} also propose a 3-layer disk model. The middle layer can form where the optical depth for the radiation reprocessed by the upper layer is less than unity. Our results suggest the existence of such a three-layer structure.


\section{Summary}
This paper presents observational results of the disk around HD163296 in $^{12}$CO ($J$=1--0), $^{12}$CO ($J$=3--2), $^{13}$CO ($J$=1--0), and $^{13}$CO ($J$=3--2) emission lines. Double-peaked profiles originated from a rotating circumstellar disk were successfully detected in all the CO lines. Physical parameters of the disk such as temperature distribution, surface density distribution and outer radius are estimated by model fitting utilizing a disk model. These physical values obtained by single dish observation were confirmed to be consistent with the results of interferometric observations \citep{Isella07}. 

The disk must be optically thick for all the CO lines. Taking advantage of difference in position of the photosphere among the CO lines, we revealed temperature distribution in vertical direction. It is proven that there are at least two distinct temperature layers, possibly three layers, from the model fitting. The temperature in the uppermost $^{12}$CO ($J$=3--2) emitting layer is estimated to be $58.5\pm 9.5$ K at 100 AU from the central star, and it is about twice higher than that in the inner regions emitting the other CO lines, $31\pm15$ K at 100 AU. Since the vertical temperature distribution inside of a disk is commonly suggested for both TTS and HAe stars (DM Tau, AB Aur, and HD31648), such a temperature distribution may be ubiquitous in protoplanetary disks \citep{Pietu07}. The temperature and density structure of protoplanetary disks are still open questions. Better knowledge such as the radial and vertical density distribution will be provided by high-resolution observations in upcoming ALMA era.


\end{document}